# Design and experimental test of an optical vortex coronagraph

Chengchao Liu [a,b], Deqing Ren [c, a, b], Yongtian Zhu [a,b], Jiangpei Dou [a,b],

a. National Astronomical Observatories/Nanjing Institute of Astronomical Optics & Technology, Chinese Academy of Sciences, Nanjing 210042,China
b. Key Laboratory of Astronomical Optics & Technology, Nanjing Institute of Astronomical Optics & Technology, Chinese Academy of Sciences, Nanjing 210042,China
c. Physics & Astronomy Department, California State University Northridge, 18111 Nordhoff Street, Northridge, California 91330-8268

## Abstract

The optical vortex coronagraph (OVC) is one of the promising ways for direct imaging exoplanets because of its small inner working angle and high throughput. This paper presents the design and laboratory demonstration performance at 633nm and 1520nm of the OVC based on liquid crystal polymers (LCP). Two LCPs has been manufactured in partnership with a commercial vendor. The OVC can deliver a good performance in laboratory test and achieve the contrast of the order $10^{-6}$ at angular distance $3\lambda/D$, which is able to image the giant exoplanets at a young stage in combination with extreme adaptive optics.

## 1 INTRODUCTION

Nowadays more than 3000 exoplanets have been discovered. Yet, most of the exoplanets have been achieved by indirect detection techniques, such as redial velocity approach or transiting method with very few exoplanets having been imaged. However, the direct imaging method acts as an important part in analyzing the atmospheric compositions of planets via spectrograph, due to its capturing photon from planets. Since the large ration between the host star and its planets as well as the small angular separation between star and planets, the direct imaging is extraordinary technological challenging. Recently, various coronagraph concepts have been proposed to suppress the central starlight; for example, amplitude apodization coronagraph (Kasdin et al. 2003; Guyon 2003), apodization pupil coronagraph (Ren&Zhu 2007; Soummer et al. 2003; Kuchner & Traub 2002) and phase mask coronagraph (Rouan et al. 2000; Abe et al 2001; Palacios 2005; Mawet et al 2005). Among all of them the vortex coronagraph has been promising because of its advantages as very small Inner Working Angle (IWA), which is one way to make using of the smaller telescope, high throughput (theoretical 100%), clear $360^{0}$ discovery region, low chromatic dependence and ease installation in coronagraphic systems (O. Guyon et al 2006).

The optical vortex coronagraph can be divided into two types: one is the scalar optical vortex coronagraph; the other is the vectorial optical vortex. With the technical breakthrough a new promising technique based on birefringent liquid crystal polymer (LCP) has been used to manufacture the optical vortex phase plate (Mawet et al. 2009a, Opt. Express). Here we report our initial design and demonstration of optical vortex coronagraph based on the rotationally symmetric half wave plate (HWP) developed by Thorlabs company.

In this paper we present our optical vortex coronagraph including the instrument design and laboratory demonstration. Theoretical description and layout of optical vortex coronagraph are shown in section 2. Laboratory measurements will be presented in section 3. Section 4 will describe our conclusions and future work.

**2 Optical vortex coronagraph design**

**2.1 Vortex coronagraph principle**

An optical vortex coronagraph uses a helical phase of the form $e^{i\phi}$, with $\phi = l\theta$, where $l$ is the topological charge and $\theta$ is the focal plane azimuthal coordinate. In optical systems vortices manifest themselves as dark donut of destructive interference that occur at phase singularities (D. Rozas et al 1997; A. Niv et al 2006). We can use this dark core to attenuate an on-axis star so nearby planets can be imaged. An optical vortex beam propagates in the z direction can be expressed by the electric field (M. Vasnetsov and K. Staliunas, et al 2002):

$$E(\rho, \phi, z, t) = A(\rho, z) \exp(il\theta) \exp(i\omega t - ikz) \qquad (1)$$

where (ρ, ϕ, z) are cylindrical coordinates, $A(\rho, z)$ is a circularly symmetric amplitude function and k=2π/λ is the wavenumber of a monochromatic field of wavelength λ. Figure 1 shows the vortex phase mask how to suppress the on-axis light.

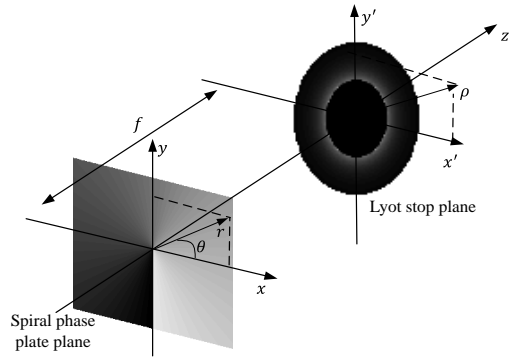

Figure 1. Illustration of the diffraction effect of vortex phase mask on the on-axis light. Most of the light will emerge outside of the geometrical circle, which we can apply a Lyot stop to block it (right bottom). FT denotes the Fourier transform.

## 2.2 Liquid crystal polymer

Different from other vortex coronagraphs such as the four-quadrant phase mask, a rotationally symmetric half wave plate manufactured by liquid crystal polymer, which can generate an azimuthal phase spiral reaching an even multiple of $2\pi$ readian in the circuit about the center is used in our test (McEldowney, S.C., 2008 OL; McEldowney, S.C., 2008 OE;). Thorlabs' LCP is a half wave retarder designed to affact the radial and azimuthal polarization of opitcal fields. A LCP offers constant retardance accross the clear aperture but its fast axis rotates continuously over the area of the optics about the center. The LCP with $l=2$ is used in our test and figure 2 shows the phase changing from 0 to $4\pi$ readian and the photograph of the LCP half waveplate.

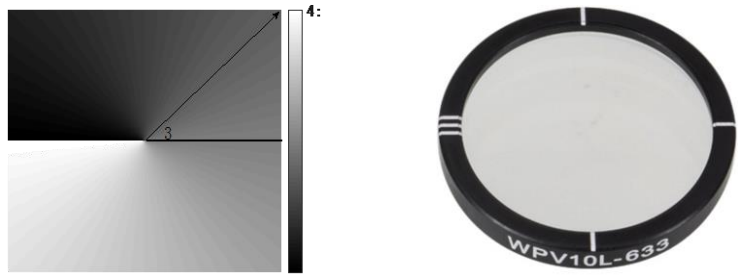

Figure. 2 The phase map of the LCP with $l=2$ (left) and the photograph optimized at 633nm.

## 2.3 Layout of the optical vortex coronagraph

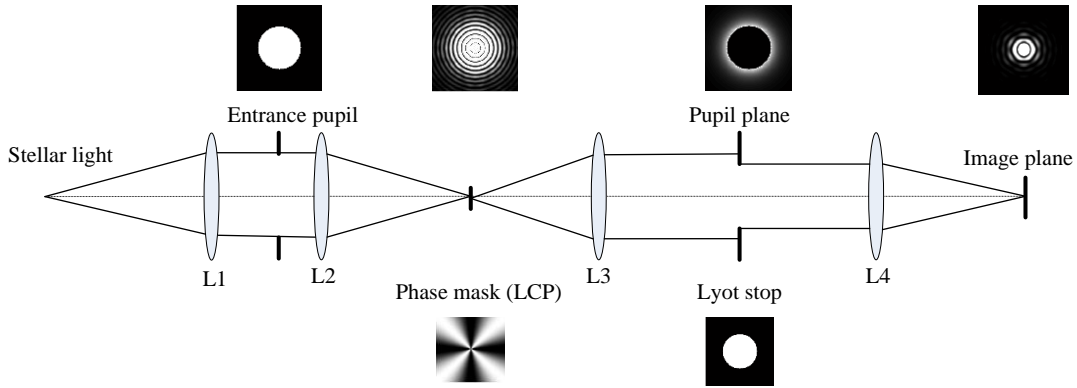

Figure 3. Schematic of optical vortex coronagraph. The stellar light simulated by laser firstly collimated by lens L1, then focused by lens L2 onto the phase mask (LCP). L3 reimages the entrance pupil where the diffraction light from the phase mask is blocked by the Lyot stop. L4 focuses the light on our camera.

A preliminary schematic diagram of OVC is shown in Fig. 3. The coronagraph includes three Fourier transforming lenses L2, L3 and L4. The entrance plane wave of the star light can be described by:

$$U(x, y)=A(x, y) \qquad (2)$$

Therefore, the electric field in the first focal plane is:

$$U(u, v)=exp(il\theta)\ FT(A(x, y)) \qquad (3)$$

The L3 performs an inverse Fourier transform on *U(x, y)* and the electric field in the pupil plane can be expressed by:

$$U(x', y') = P_{Lyot}(x', y')FT^{-}(exp(il\theta)FT(A(x, y)))) \quad (4)$$

$P_{Lyot}$ is the Lyot stop transmission function expressed by $P_{Lyot}(x', y') = 1$ for $\rho = \sqrt{x'^2 + y'^2} < R_{Lyot}$ and zero otherwise. The electric field on the final image plane can be described as:

$$U(u', v')=FT(U(x', y')) \quad (5)$$

We simulated the performance of the OVC according to above equations. In order to obtain high sampling of the coronagraphic images, 4096 x 4096 pixel matrices are used. In our simulations, we consider a ideal plane wavefront entering an unobstructed telescope with a circular aperture with diameter D=200 pixels. The Lyot stop diameter $D_{LS}$ is set to 0.8 times that of the entrance pupil. The topological charge of LCP is $l = 2$. Figure 4 describes the simulation Lyot stop image after the phase mask and the contrast achieved by the OVC with the *l*=2. From the figure we can conclude that the theretical contrast with a vortex phase mask (*l*=2) can reach $10^{-7}$ at $2\lambda/D$ (blue line) comparison to non-coronagraphic contrast (red line). Due to the vortex coronagraph sensitivity to the finte pixels sampling which constraints the numercial simulation, it can not achieve the theoretical 100% nulling. Therefore, the contrast 10-7 is a reasonable theoretical contrast value with a reasonable number pixels for sampling.

## 3 OVC laboratory measurements

We tested the performance of the OVC using two different LCP phase masks at wavelength 633nm and 1520nm. Then the measurement contrast result is disscused.

### 3.1 Optical setup

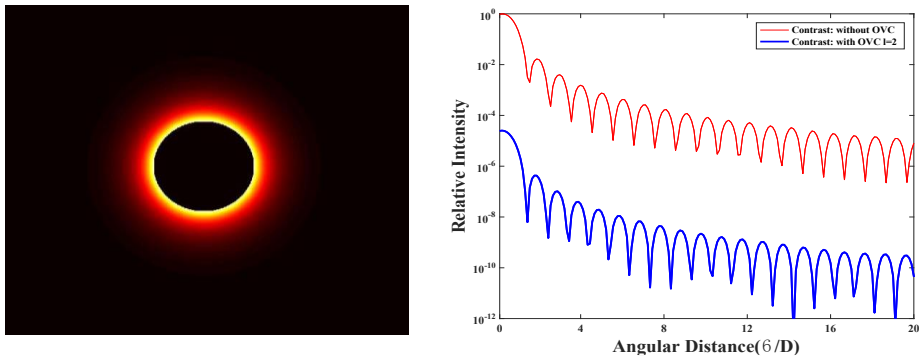

Figure 4. The theoretical Lyot stop image (left) illustrates the on-axis light is blocked. The theroetical contrast can reach to the order of $10^{-7}$ at the angular seperation $2\lambda/D$ (right).

Figure 3 shows the layout of the OVC optical system, together with the phase mask, point spread function (PSF) on different focal plane and a Lyot stop. Both of the starlight and planet light are simulated by two point sources using two laser light soures. A diaphram is used as the entrance pupil and the collimated light is focused on the phase mask. It is a critical issue to adjust the focus to the center of the phase mask exactly. Therefore the LCP is installed on a three dimensional adjusting holder, which allows fine translations to locate its center perfectly at the focal point. A rotationally symmetric half wave plate generates the phase spiral in vector vortex mask. Two liquid crystal polymers optimized at 633nm and 1520nm respectively are manufactured by Thorlabs. The transimission of the LCPs is better than 97%. The Lyot stop is used to block the light outside the pupil and is conjugated with the entrance pupil through our optical system. The diameter of the entrance pupil and the Lyot stop is 10mm and 8mm respectively. The detectors used in this test are 16-bit SBIG and 14-bit near infrared Xenics camrera.

### 3.2 Image acquisition and data reduction

To estimate the peformance of OVC at 633nm and 1520nm, we first recoded 100～1000 images with the specific exposure time. The number of farames is choosen to reduce the read-out noise. To avoid the saturation of noncoronagraphic images we used netural density filters and shorter exposures. The netural density filters have been calibrated with the detector and could relize an attenuation factor of $10^{-3}$. Then we removed the LCP phase mask to grab non-coronagraphic images with the same exposure time. Also all the images are subtracted with a median dark frame (derived from 100 individual darks). In this test the peak to peak rejection metric, which means to calculate the ration of the maximum intensity of the PSF without phase mask to the maximum intensity of the PSF with coronagraph, is adopted to qualify the OVC performance to attenuate starlight.

### 3.3 Laboratory test results

We first grabed an image of the coronagraphic exit pupil to verify its performance is consistent with theoretical simulation. Figure 5a shows the exit pupil and the stellar light is mostly rejected outside of the pupil geometical cirlce with the diffraction pattern. A Lyot stop is deployed at the exit pupil plane to block the stellar light. Figure 5b presents the PSF without phase mask apodization in the final focal plane, where the planet is submerged in the stellar light diffration. The PSF with the phase mask is shown in Figure 5c, in which the planet appears at the angular diastance $2\lambda/D$ with the contrast of $10^{-4}$. The attenuation ration of the PSF peak with the OVC to the PSF peak without the OVC is about $3.03 \times 10^{-4}$. Figure 6 shows that our optical vortex coronagraph can achieve the contrast of $10^{-6}$ at small angular distance $3\lambda/D$ both at visible(633nm) and infrared wavelength(1520nm).

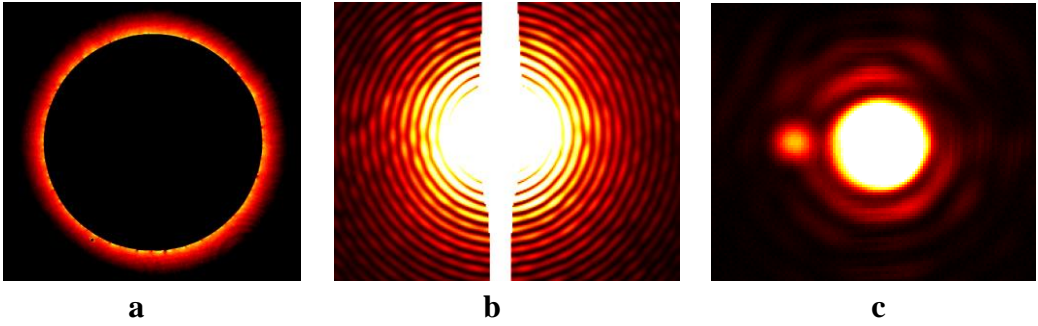

Figure 5. The exit pupil after the phase mask (a), the PSF without phase mask (b) and the PSF with the phase mask (c).

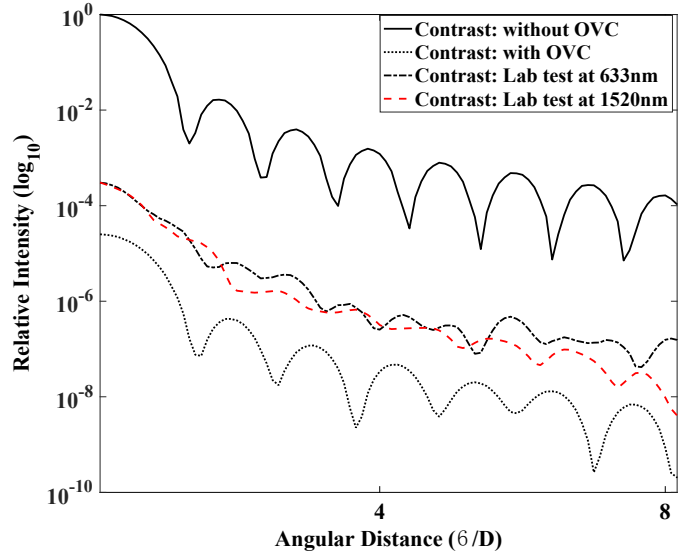

Figure 6. Laboratory contrast test of our visible (dash-dotted line) and infrared (red dashed line) optical vortex corongraph.

## 4. Conclusions and future work

Considering the high throughput, small inner working angle and complete 360º discovery space, the optical vortex coronagraph has been a promising technique for direct imaging exoplanets using small telescopes, which have enough observation time for high contrast imaging exoplanets comparison to large telescopes. In this paper we propose our optical vortex coronagraph based on the liquid crystal polymer including the numerical simulation and laboratory experiment demenstration. Both the visible and infrared vortex phase masks are used in the laboratory test. The simulation result shows that the OVC can achieve the contrast of $10^{-7}$ at $2\lambda/D$. The laboratory test using visible and infrared LCPs manufactured by Thorlabs demenstrates that a deep suppression of the star peak (3300) can be reached. The contrast of $10^{-6}$ at angular distance $3\lambda/D$ has been realized at visible and infrared wavelength.

Next stage this optical vortex coronagraph can be integrated with our extreme adaptive optics (Ex-AO) to direct image the young gaint exoplanets with the contrast around $10^{-6}$ (Ren et al; Dou et al 2015). Also our mature stepped-transmission coronagraph technique can adopt this vortex phase mask to compose a hybrid coronagraph to obtain better performance. New progress will be presented in the following papers.

**Acknowledgements**

This work is partly supported by Mt. Cuba Astronomical Foundation.